\documentclass[twocolumn,aps,prl,floatfix,superscriptaddress]{revtex4-1}
\usepackage{amsmath,amssymb}
\usepackage{graphicx,float}
\usepackage{times,txfonts}
\usepackage{textcomp}
\usepackage{color}
\usepackage{multirow}
\usepackage{color}
\usepackage{soul}
\usepackage{hyperref}
\usepackage{natbib}
\usepackage{nicefrac}
\usepackage{multirow}
\usepackage{blindtext}
\usepackage[export]{adjustbox}
\definecolor{hugoColor}{RGB}{59,134,255}

\begin{document}

\title{Universal linear optics by programmable multimode interference}

\author{Hugo Larocque}
\affiliation{Research Laboratory of Electronics \& Department of Electrical Engineering and Computer Science, Massachusetts Institute of Technology, Cambridge, Massachusetts 02139, USA}

\author{Dirk Englund}
\affiliation{Research Laboratory of Electronics \& Department of Electrical Engineering and Computer Science, Massachusetts Institute of Technology, Cambridge, Massachusetts 02139, USA}
\email{hlarocqu@mit.edu}

\begin{abstract}

We introduce a constructive algorithm for universal linear electromagnetic transformations between the $N$ input and $N$ output modes of a dielectric slab. The approach uses out-of-plane phase modulation programmed down to $N^2$ degrees of freedom. The total area of these modulators equals that of the entire slab: our scheme satisfies the minimum area constraint for programmable linear optical transformations. We also present error correction schemes that enable high-fidelity unitary transformations at large $N$. This ``programmable multimode interferometer'' (ProMMI) thus translates the algorithmic simplicity of Mach-Zehnder meshes into a holographically programmed slab, yielding DoF-limited compactness and error tolerance while eliminating the dominant sidewall-related optical losses and directional-coupler-related patterning challenges. 

\end{abstract}

\maketitle

\section{Introduction}

\begin{figure}[t!]
	\centering
	\includegraphics[width=\linewidth]{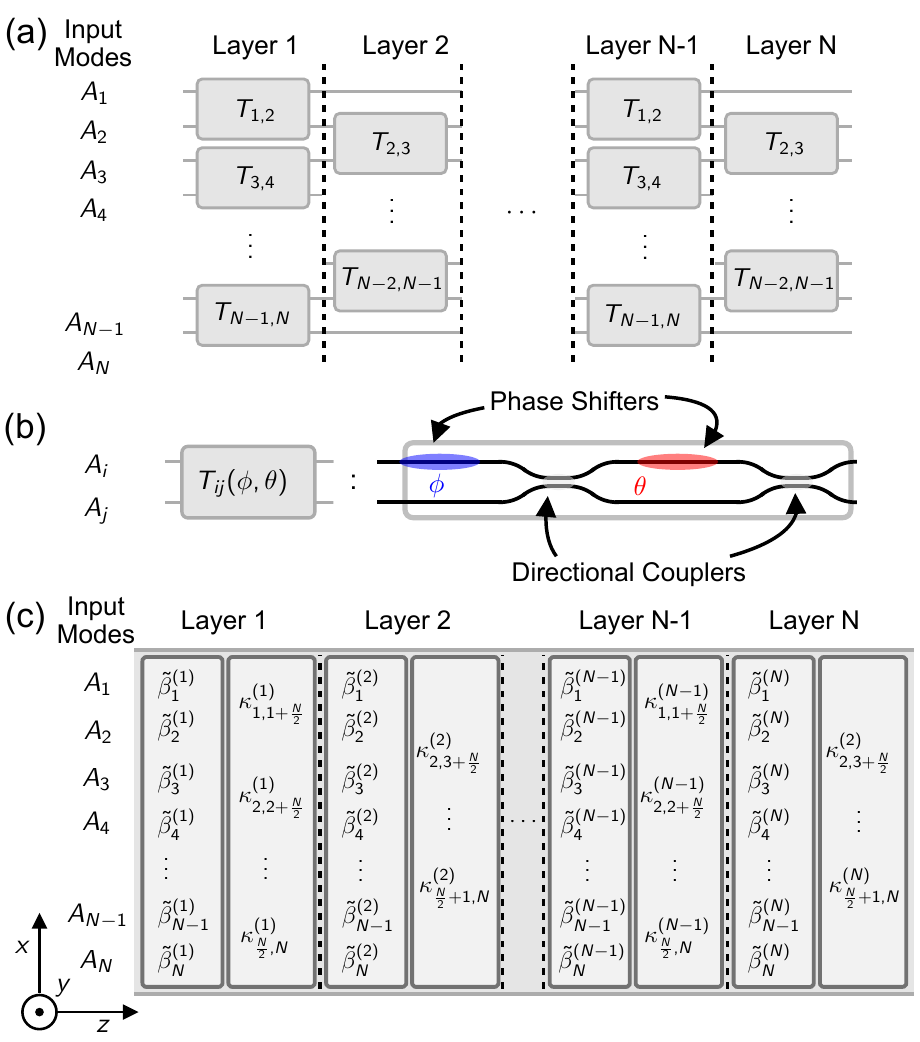}
	\caption{\textbf{SU(2) mesh architectures} (a) Schematics of an SU(N) transformation decomposed into a rectangular mesh of SU(2) transformations $T_{ij}$. (b) Mach-Zehnder interferometer (MZI) used in integrated photonic platforms for implementing individual $T_{ij}$s in the mesh. These building blocks consist of two phase shifters, where one of them is book-ended by directional couplers. (c) MMI approach to realizing SU(N) transformations. The latter are achieved by segmenting the waveguide into layers where programmed refractive index variations first alter the propagation constants $\beta_{m,n}$ of modes $m,n$ to $\tilde \beta^{(l)}_{m,n}$ and subsequently couples them as parametrized by the coupling strength $\kappa^{(l)}_{mn}$.}
	\label{fig:figConcept}
\end{figure}

Rapid advances in photonic integrated circuits (PICs) have ushered in a new generation of devices capable of universal linear-optics transformations across a set of $N$ optical modes~\cite{Bogaerts:20}. These `programmable photonic circuits' are enabling new applications ranging from quantum information processing~\cite{Harris:18,Rudolph:17} to deep learning~\cite{Wetzstein:20}. In leading implementations, the mode transformations are realized as an SU($N$) rotation, which can be factored into $N(N-1)/2$ SU(2) rotations~\cite{Murnaghan:58}. Fig.~\ref{fig:figConcept}(a) illustrates the concept on input modes $A_1, ..., A_N$, where SU(2) rotations are programmed in Mach-Zehnder interferometers (MZIs), as shown in Fig.~\ref{fig:figConcept}(b). A key advantage of this `MZI mesh' (MZM) architecture is its compatibility with efficient programming algorithms~\cite{Reck:94,Clements:16}.  But while tens of modes have been realized~\cite{Bogaerts:20}, the MZM architecture has inherent limitations -- most importantly, the exacting fabrication requirements of 50:50 couplers, a large fraction of inter-waveguide `dead' space, and architecture-constrained operating bandwidth and decomposition schemes~\cite{Simon:17, deGuise:18}.

Alternatively, could universal linear-optical transformations be deterministically programmed in a rectangular dielectric slab, a large `multimode interferometer'? We consider this question for the $N$ transverse spatial modes of a slab with an index contrast quantified by$\sqrt{\Delta \epsilon/\epsilon_0}$, as shown in Fig.~\ref{fig:figConcept}(c). This slab has single-mode thickness $L_y= \lambda/2 \sqrt{\Delta \epsilon/\epsilon_0}$, length $L_z$, while the width $L_x$ accommodates $N\sim 2 L_x \sqrt{\Delta \epsilon/\epsilon_0}/\lambda $ transverse spatial modes. Recent experimental work did realize programmable $1\times 2$ switching in a 2-mode MMI fabricated in silicon~\cite{Bruck:16}, but programming the index perturbations (applied via photogenerated carriers) relied on a randomized search algorithm that lacks the algorithmic determinism of the MZM approach ~\cite{Reck:94, Clements:16, Miller:17}. On the other hand, a recent theoretical proposal did find a deterministic algorithm for programming SU($N$) transformations~\cite{vanNiekerk:19} by cascading $\mathcal{O}(N)$ layers of phase shifters and self-imaging planes, each of length $\mathcal{O}(N)$. But, this device's length scales as $L_z \propto N^2$, whereas the minimum necessary scaling is $L_z\propto N$ to accommodate the necessary $N(N-1)$ phase shifters.

Here, we show that the compactness of MMIs and the programmability and $\mathcal{O}(N)$ length scaling of MZMs are possible in one device: a programmable multimode interferometer (ProMMI) that relies on a mesh of $\mathcal{O}(N^2)$ phase shifters distributed across a rectangular waveguide slab. Waveguide sidewall scattering loss is largely eliminated in the large cross-section waveguide; transverse dimensions are reduced by $\sim\! 1$ order of magnitude compared to MZMs; fabrication challenges of $2\times 2$ couplers are eliminated. The longitudinal dimension scales as $L_z=\text{max}(\alpha_1 N,\alpha_2 N^2)$, where $\alpha_1,\alpha_2$ are coefficients that depend on the ProMMI material platform. We also translate recent hardware error correction methods from MZM~\cite{Bandyopadhyay:21} to  ProMMI architectures, allowing substantial improvements in unitary transformation fidelity.

\section{Results}

Instead of operating on an array of meshed single mode waveguides, the ProMMI architecture implements SU(N) transformations on the eigenmodes of a slab. Here, the slab has a width $L_x=w$, allowing it to support $N$ modes. As in the MZM scheme~\cite{Clements:16}, we cascade these transformations into $N$ layers performing SU(2) transformations on pairs of eigenmodes. But unlike the MZM, we apply these transformations via structured index perturbations that first modify a mode's propagation constant from $\beta_m$ to $\tilde\beta_m$ and then couple modes $m$ and $n$ at a rate $\kappa_{mn}$. Each $(\tilde\beta_{m,n}, \kappa_{mn})$ has two DoFs as required to control, say, the independent variables of relative phase and amplitude. But, unlike MZMs, the ProMMI architecture overlays these perturbations. In this way, the ProMMI achieves the deterministic programmability and length scaling of the MZM, while collapsing its width to $w$, i.e., the minimum required for $N$ transverse modes.

\vspace{10pt}
\noindent \textbf{Coupled Mode Theory.} 
We approximate the step-index slab as infinitely thick such that its permittivity $\epsilon_s(\mathbf{r})=\epsilon_\text{core}$ for $|x|<w/2$ and $\epsilon_s(\mathbf{r})=\epsilon_\text{clad}$ elsewhere. The monochromatic transverse electric (TE) field propagating along the waveguide in the $z$ direction thus takes the form 
\begin{equation}
    \mathbf{E}(x,y,z) = u(x,y,z) e^{i\omega t} \hat{\mathbf{y}},
\end{equation}
where the $\hat{\mathbf{y}}$ component of the field, $u(\mathbf{r})$, can be expanded in the orthonormal TE eigenmode basis $\{A_n\}$ as
\begin{equation}
   u(x,y,z) = \sum_n a_n(z) A_n(x,y) e^{-i \beta_n z }.
\end{equation}
We use coupled mode theory~\cite{Yariv:73} to find the permittivity perturbation of the form $\epsilon^{(l)}_{p,(m,n)}(\mathbf{r})=\epsilon_{A,(m,n)}^{(l)}\epsilon_{L,(m,n)}^{(l)}(z)\epsilon_{T,(m,n)}^{(l)}(x,y)$ for the SU(2) rotation between modes $m$ and $n$ in the $l^\text{th}$ layer of the ProMMI. Substituting $\mathbf{E}(x,y,z)$ into the wave equation
\begin{equation}
    \boldsymbol{\nabla} \times \boldsymbol{\nabla} \times \mathbf{E} = \frac{\epsilon_s(\mathbf{r})+\epsilon^{(l)}_{p,(m,n)}(\mathbf{r})}{c^2} \frac{\partial^2 \mathbf{E}}{\partial t^2}.
\end{equation}
yields the coupled mode equations
\begin{equation}
    \label{eq:mismatch}
    \frac{\partial{a_m}}{\partial z} e^{-i\beta_m z} = \frac{1}{\beta_m} \sum_n a_n \epsilon_{A,(m,n)}^{(l)}\epsilon_{L,(m,n)}^{(l)}(z)\kappa^{(l)}_{mn} e^{-i\beta_n z},
\end{equation}
where $\kappa_{mn}^{(l)}$ is the coupling coefficient:
\begin{equation}
    \label{eq:overlap}
   \kappa^{(l)}_{mn} = -\frac{i}{2} \frac{\omega^2}{c^2} \int A_m^*(x,y) \epsilon_{T,(m,n)}^{(l)} (x,y) A_n(x,y) \, dx \, dy.
\end{equation}

To optimize the selective coupling between modes $m$ and $n$, we therefore set $\epsilon_{L,(m,n)}^{(l)}(z)= \cos((\beta_m-\beta_n)z)$ and use Eq.~(\ref{eq:overlap}) to determine $\epsilon_{T,(m,n)}^{(l)}(x,y)$ that maximizes $\kappa^{(l)}_{mn}$.
\begin{figure*}[t!]
	\centering
	\includegraphics[width=1\linewidth]{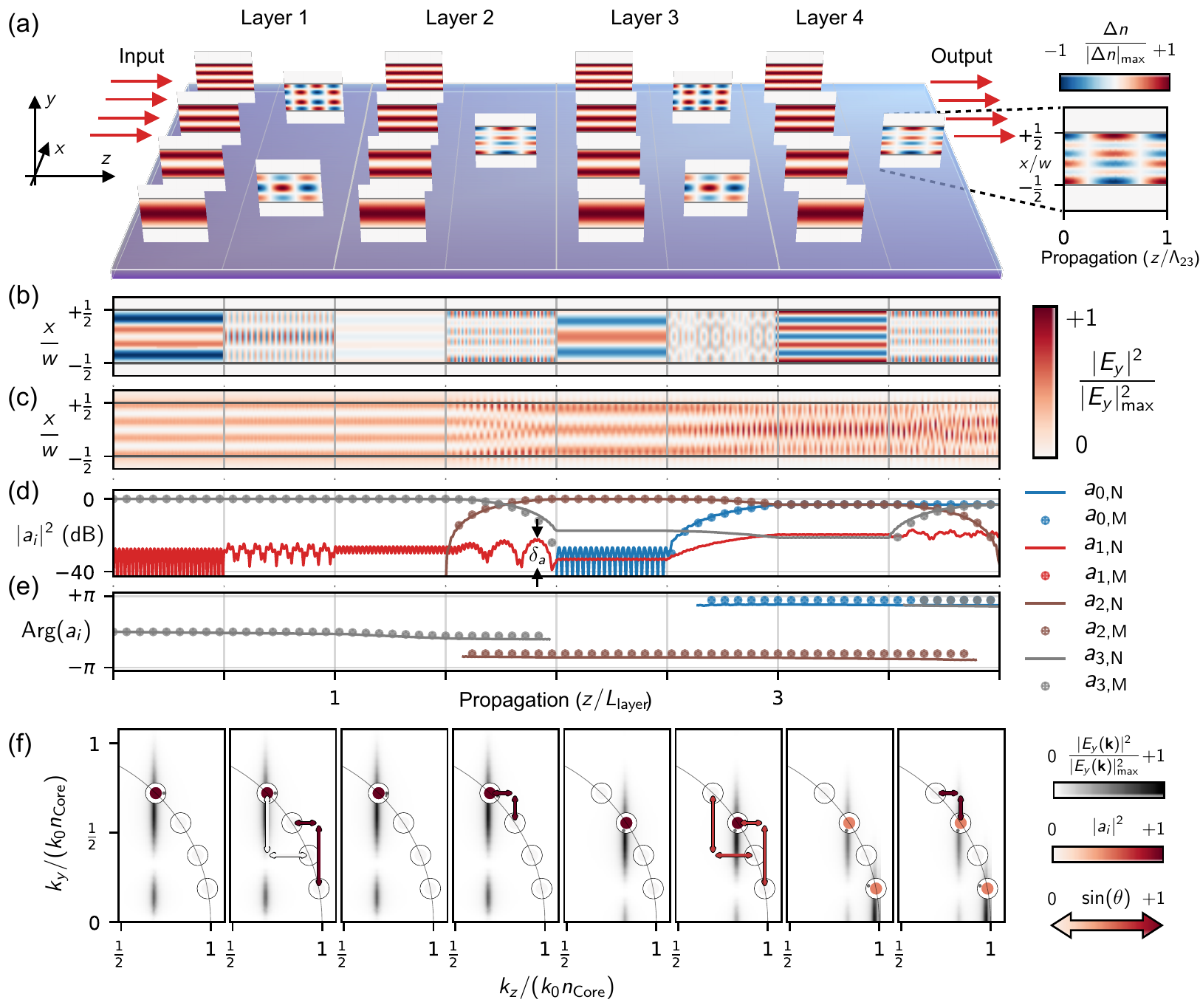}
	\caption{\textbf{ProMMI architecture of a 4-mode slab waveguide.} (a) Structured refractive index perturbations used to realize arbitrary SU(4) transformations. The location of each perturbation matches that of the parameter it modifies according to Fig.~\ref{fig:figConcept}(c). Inset: refractive index perturbation used to couple the waveguide's TE2 mode to the TE3 mode, where $\Lambda_{23}=2\pi/(\beta_2-\beta_3)$. (b) Refractive index perturbations required to induce the SU(4) transformation provided in Eq.~(\ref{eq:su4}). (c) Intensity of the optical field upon propagating through the perturbed waveguide for an input mode defined as $(a_1,a_3,a_2,a_0)^T=(0,1,0,0)^T$, which in our architecture, consists of the waveguide’s TE3 mode. Refractive indices of $n_\text{core}=2.2886$ and $n_\text{clad}=1.44$ were assigned to the core and the cladding of the waveguide, respectively, and we considered an optical wavelength of 1.55~\textmu m. (d) Absolute square and (e) phase values of the mode expansion coefficients, $a_{i,\text{N}}$, of the simulated field in (c) overlaid onto circular markers plotting the coefficients, $a_{i,M}$, expected from Eq.~(\ref{eq:su2}). (f) Optical transformations in each layer represented in $k$-space. The density plots show the absolute square of the field's Fourier transform at the start of the phase shifting and coupling sections of each layer. Markers located at each mode's central $\mathbf{k}$ value are colored based on their weight and the angular position of the dot on their outline corresponds to their phase. Vertical and horizontal arrows indicate the propagation constant shifts enabled by the $\epsilon_{L,(m,n)}^{(l)}$ and $\epsilon_{T,(m,n)}^{(l)}$ components of the permittivity perturbations $\epsilon_{p,(m,n)}^{(l)}$, respectively, and are colored based on their corresponding $\theta$ values.}
	\label{fig:figArchitecture}
\end{figure*}
This $\epsilon^{(l)}_{p,(m,n)}(\mathbf{r})$ in principle allows coupling between other pairs of modes $m',n'$ as indicated by the maximum transfer amplitude
\begin{eqnarray}
    \label{eq:maxTransfer}
    \eta_{m'n'} &=& \frac{2 |\kappa^{(l)}_{m'n'}|}{\sqrt{4 |\kappa^{(l)}_{m'n'}|^2+(\beta_p - \left(\beta_{m'}-\beta_{n'})\right)^2}}\\ \nonumber
    &=& \frac{1}{\sqrt{1+\left(\frac{\beta_p - (\beta_{m'}-\beta_{n'})}{2|\kappa^{(l)}_{m'n'}|}\right)^2 }},
\end{eqnarray}
where $\beta_p=\beta_m-\beta_n$. Therefore, this cross-talk is  suppressed when 
\begin{equation}
    |\kappa^{(l)}_{m'n'}|/|(\beta_m-\beta_n)-(\beta_{m'}-\beta_{n'})| \ll 1,
\end{equation}
which can be achieved in practice by lowering $\epsilon_{A,(m,n)}^{(l)}$.

We set the propagation constants $\tilde{\beta}_m^{(l)}$ in layer $l$ (Fig.~\ref{fig:figConcept}(c)) with the transverse perturbations $\epsilon_{T,(m,n)}^{(l)}(x,y)$ while turning off the mode coupling term, i.e., $\epsilon_{L,(m,n)}^{(l)}(z)=1$. Across the full layer $l$ of length $L_\text{layer}$, the perturbation $\epsilon^{(l)}_{p,(m,n)}(\mathbf{r})$ thus transforms modes $m$ and $n$ as 
\begin{equation}
    \label{eq:su2}
    T^{(l)}_{mn} = 
    \begin{pmatrix}
        e^{-i\phi} \cos{\theta} & -i \sin{\theta} \\
       -i e^{-i\phi}\sin{\theta} & \cos{\theta} \\
    \end{pmatrix},
\end{equation}
where $\theta= \kappa^{(l)}_{mn}L_\text{layer}/2$ is set by the mode coupler and $\phi = ((\tilde\beta^{(l)}_m-\tilde\beta^{(l)}_n)-(\beta^{(l)}_m-\beta^{(l)}_n))L_\text{layer}/2$ by the phase shifting perturbations. This $m,n$ SU(2) transformation is the basic building block of our SU(N) decomposition algorithm, which is similar to the Clements scheme~\cite{Clements:16}, as detailed in the Supplementary. 

By this procedure, we determine all the necessary permittivity perturbations $\epsilon_{p,(m,n)}^{(l)}$ to construct the total perturbation  $\epsilon_{p}^{(l)}(\mathbf{r}) = \sum_{(m,n)}\epsilon_{p,(m,n)}^{(l)}(\mathbf{r})$ in layer $l$. We repeat the process for all $l$ to complete the desired SU(N) on the slab's eigenmodes.

\vspace{10pt}
\noindent \textbf{4-mode Transformation.} Let's consider an exemplary unitary transformation SU(4) on the TE modes of a $N=4$ slab waveguide, with matrix elements: 
\begin{equation}
    \label{eq:su4}
    \frac{1}{\sqrt{2}}
    \begin{pmatrix}
        -i & 0 & 0 & 1 \\
        0 & -i & 1 & 0 \\
        i & 0 & 0 & 1 \\
        0 & i & 1 & 0
    \end{pmatrix}.
\end{equation}
Fig.~\ref{fig:figArchitecture}(a) shows the $T_{mn}^{(l)}$ building blocks of each layer: the refractive index shifts $\Delta n$ corresponding to four phase-shift perturbations with $\epsilon_{L,(m,n)}^{(l)}(z)=1$ and the coupling perturbations with $\epsilon_{L,(m,n)}^{(l)}(z)=\cos((\beta_m-\beta_n)z)$.

Fig.~\ref{fig:figArchitecture}(b) shows the index shift caused by the total perturbation $\sum_{(m,n),l}\epsilon^{(l)}_{p,(m,n)}(\mathbf{r})$ implementing Eq.~(\ref{eq:su4}). 
Figure~\ref{fig:figArchitecture}(c) plots the $|E_y(x,0,z)|^2$ of a trial input field $(a_1,a_3,a_2,a_0)^T=(0,1,0,0)^T$. Here, we used  a split-step Fourier method~\cite{Feit:78} to numerically propagate the optical fields. As expected, $|E_y|^2$ is flat during the first half of layer 1 given that the only eigenmode propagating in the waveguide experiences a global phase shift. No coupling occurs in the second half because $\epsilon_{p,(3,n)}^{(1)}(\mathbf{r})=0$. In the second layer, the field remains flat in the first half for the same reasons as it did in layer 1, whereas it is entirely converted to $A_2$ in the second half since $\kappa^{(2)}_{2,3}L_\text{layer}/2=\pi/2$. The first half of $l=3$ features a mostly flat field as in $l=1,2$ and partial conversion to $A_0$ as specified by $\kappa^{(3)}_{0,2}L_\text{layer}/2=\pi/4$. The relative phase between $A_0$ and $A_2$ manifests in the beating pattern in the first half of $l=4$ followed by the full conversion of $A_2$ to $A_3$ according to $\kappa^{(4)}_{2,3}L_\text{layer}/2=\pi/2$ in the second half.
In Fig.~\ref{fig:figArchitecture}(d), (e), we provide the expansion coefficients of the field and compare them to those obtained directly from Eq.~(\ref{eq:su2}). 

For further insight, it helps to view the transformation problem in $k$-space.   Fig.~\ref{fig:figArchitecture}(f) shows the Fourier transform of the field in Fig.~\ref{fig:figArchitecture}(c) as it enters the phase-shifting and coupling sections of each layer. Markers located at the central $\mathbf{k}$ value of each eigenstate represent their relative weights. The perturbations $\epsilon_{p,(m,n)}^{(l)}$ couple modes $m,n$ at a rate $\kappa_{mn}^{(l)}$ --- they are therefore the `$k$-space' analog of the MZI SU(2) rotations. Thus, the ProMMI is the `$k$-space' realization of an MZM programmable unitary that, however, is inherently more compact and does not need single-mode patterning.

\begin{figure}[t!]
	\centering
	\includegraphics[width=\linewidth]{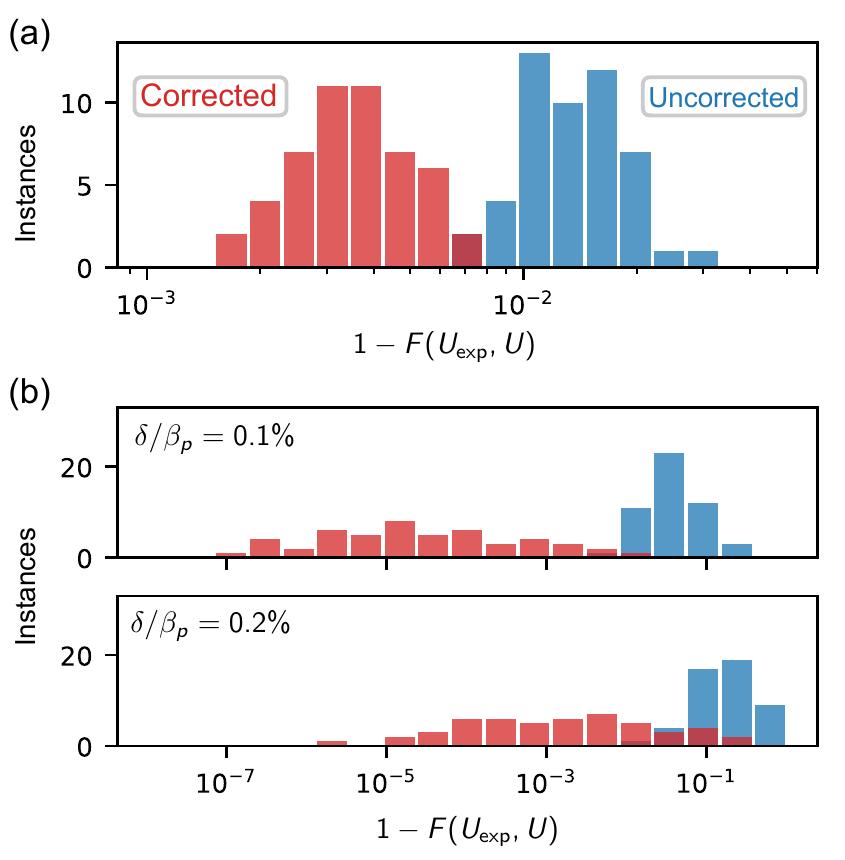}
	\caption{\textbf{ProMMI error correction.} \textbf{a,} Device infidelity $1-F(U_\text{exp}, U)$ before and after correcting for cross-talk due to partial phase matching with gradient descent optimization. \textbf{b,} Infidelity in the unitaries constructed from phase mismatched versions of Eq.~(\ref{eq:su2}) before and after correcting for imperfect phase matching. To emulate the effect of variations in the waveguide width from one layer to another, each SU(2) considered in the unitary's reconstruction was affected by a mismatch $\delta$ drawn from a normal distribution with a standard deviation of $\delta = \alpha \beta_p$. For the waveguide parameters considered in Fig.~\ref{fig:figArchitecture}, $\alpha$ values of 0.1\% and 0.2\% correspond to width discrepancies of 1.5~nm and 3~nm, respectively.}
	\label{fig:figError}
\end{figure}

\vspace{10pt}
\noindent \textbf{Error Correction.} To further gauge the reliability of our architecture, we performed the mode propagation simulations shown in Fig.~\ref{fig:figArchitecture} for 50 Haar random SU(4) rotations, $U$. We extracted the corresponding transmission matrix of the ProMMI from these simulations, $U_\text{exp}$, and thereafter the fidelity of the transformation, $F(U_\text{exp},U) = |\text{Tr}(U^\dagger U_\text{exp})/(N \, \text{Tr}(U_\text{exp}^\dagger U_\text{exp}))^{1/2}|^2$~\cite{Clements:16}. Limits in the fidelity of these transformations arise from perturbations $\epsilon_{p,(m,n)}^{(l)}(\mathbf{r})$ being partially phased-matched to couple modes $m'\neq m$ and $n' \neq n$. These conditions increase the maximum transfer amplitude $\eta_{m'n'}$ defined in Eq.~(\ref{eq:maxTransfer}), causing transformations $T_{mn}^{(l)}$ to deviate from the block-diagonal form in Eq.~(\ref{eq:su2}). For instance, $\eta_{23}\sim 0.005$ in the second half of layer 2 from Fig.~\ref{fig:figArchitecture}, explaining the oscillation in $|a_1|^2$  (see  Fig.~\ref{fig:figArchitecture}(d)) with an amplitude $\delta_a$ of roughly 0.5\% of $|a_2|^2$ in that region of the ProMMI. The cross-talk introduced by this process consequently weakens the validity of our proposed SU(N) decomposition. However, as shown in Fig.~\ref{fig:figError}(a), we can use additional optimization methods to further increase the fidelity of these transformations by altering the relative strengths of $\epsilon_{p,(m,n)}^{(l)}$. The cross-talk can also be mitigated by globally decreasing the strengths of $\epsilon_{p,(m,n)}^{(l)}$, albeit at the expense of increasing the total length of the device.

As in the case of MZMs, hardware errors can further reduce fidelity. For MZMs, the dominant source of errors is faulty couplers that do not have a 50:50 splitting ratio. For the ProMMI, such errors occur when a perturbation $\epsilon_{p,(m,n)}^{(l)}(\mathbf{r})$ has a period $\Lambda_p \neq 2\pi/|\beta_m - \beta_n|$ due to fabrication errors in the MMI waveguide that shift $\beta_{m,n}$ from their target values. In this event, the coupling between modes $m$ and $n$ is not perfectly phase-matched, which increases the coupling rate between these modes and constrains the off-diagonal terms of $T_{mn}^{(l)}$ to absolute values below unity. For instance, under these mismatched conditions, the bottom left element of $T_{mn}^{(l)}$ becomes 
\begin{equation}
    (T_{mn}^{(l)})_{21} \rightarrow -i \frac{\kappa_{mn}^{(l)}e^{i (\delta L_\text{layer}/2-\phi)}}{\sqrt{\delta^2+{(\kappa_{mn}^{(l)})}^2}} \sin\left(\sqrt{\delta^2+{(\kappa_{mn}^{(l)})}^2} \frac{L_\text{layer}}{2}\right),
\end{equation}
where $\delta= (\beta_m - \beta_n - \beta_p)/2$ and $\beta_p=2\pi/\Lambda_p$.
This modification reduces the validity of the $\theta$ and $\phi$ parameters obtained in a unitary's SU(2) decomposition. Though schemes such as gradient descent methods~\cite{Pai:19} could correct these errors, a pre-characterization of the ProMMI's perturbations can provide a more deterministic form of correction. Based on a similar scheme used for MZM hardware~\cite{Bandyopadhyay:21}, we deduce that hardware error correction in ProMMIs first involve modifying $\kappa_{mn}^{(l)}$ to ${\kappa_{mn}^{(l)}}' = ((\kappa_{mn}^{(l)})^2 + \Delta \kappa^2)^{1/2}$, where to first order
\begin{equation}
    \Delta \kappa^2 =  \delta^2 \left(\frac{\tan(\kappa_{mn}^{(l)} z)}{\kappa_{mn}^{(l)} z } - 1\right).
\end{equation}
As shown in Fig.~\ref{fig:figError}(b), letting $\kappa_{mn}^{(l)} \rightarrow {\kappa_{mn}^{(l)}}'$ with an additional modification in the $\phi$ parameter of Eq.~(\ref{eq:su2}) corrects the faulty mode couplers in the unitaries considered in Fig.~\ref{fig:figError}(a). As shown in the Supplementary, $\delta$ is related to errors in waveguide width, $\Delta w$, and perturbation period, $\Delta \Lambda$, through $\delta \approx (\beta_m-\beta_n)(\Delta w/w - \Delta \Lambda/2\Lambda)$. As a result, errors in the MMI waveguide are suppressed in devices with more modes.

\begin{figure}[t!]
	\centering
	\includegraphics[width=\linewidth]{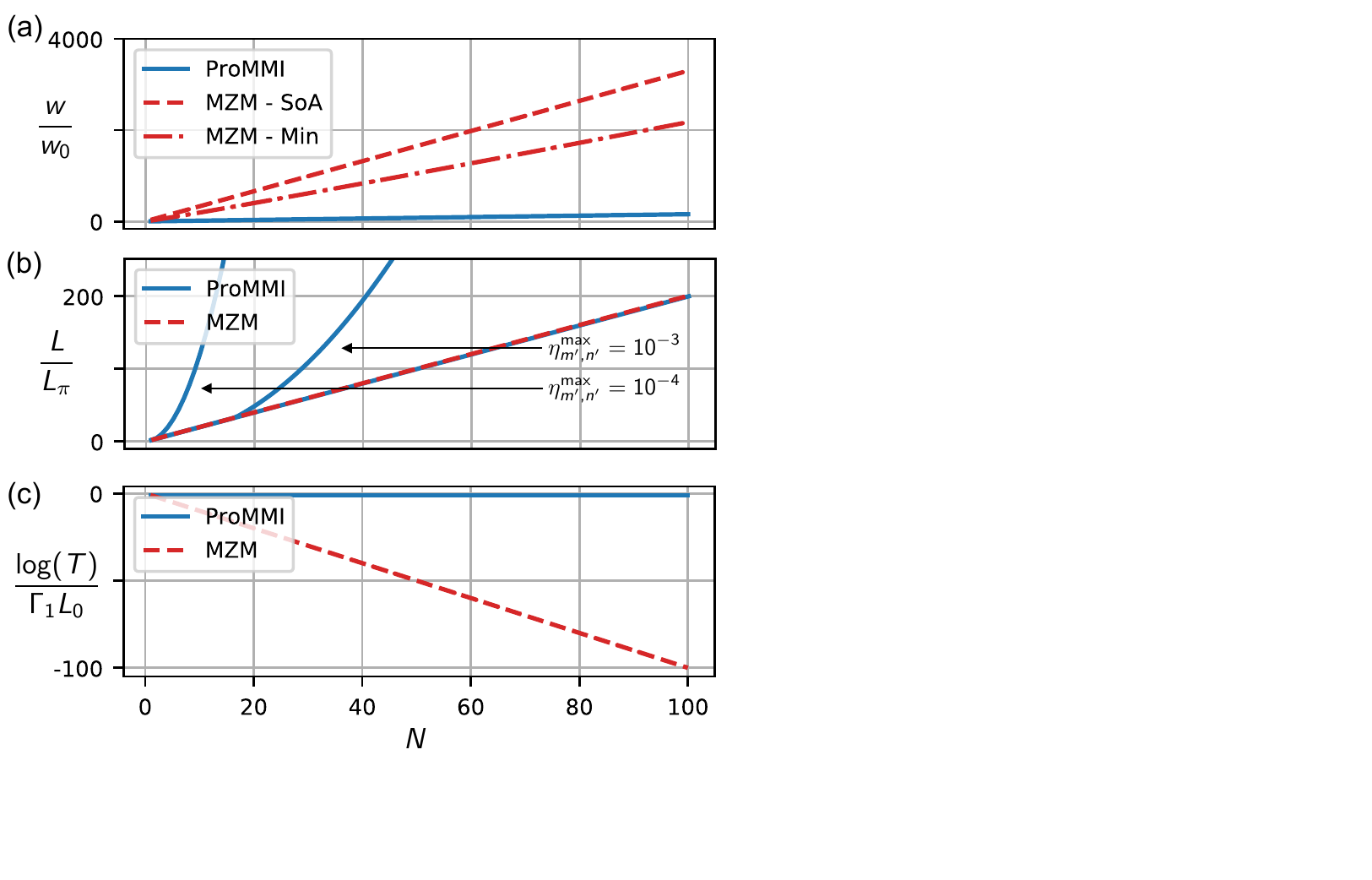}
	\caption{\textbf{Scaling properties of ProMMI vs MZM devices.} \textbf{a,} Device width in units of $w_0 = \lambda/\pi\sqrt{(n_\text{core}^2-n_\text{clad}^2)}$. \textbf{b,} Device length for both implementations in units of $L_\pi$. \textbf{c,} Optical transmission expressed in units of $\Gamma_1 L_0$, where $\Gamma_1$ is the loss rate attributed to a single mode waveguide and $L_0$ is the layer length of the mode transformer. State-of-the art (SoA) values are attributed to SOI platforms equipped with thermo-optic phase shifters. The minimum width scaling in (\textbf{a}) is attributed to the waveguide spacing that leads to less than 1\% cross-coupling between the waveguides of an MZM over its entire length.}
	\label{fig:figScaling}
\end{figure}

\vspace{10pt}
\noindent \textbf{Scaling.} The ProMMI dimensions scale favorably compared to the MZM. First, in both approaches, the device width scales linearly with $N$, $w=\alpha_{\text{MZM,ProMMI}} N$, as seen in Fig. \ref{fig:figScaling}(a). However, $\alpha_{\text{MZM}}\gg \alpha_{\text{ProMMI}}$ to allow for spacing between the MZM waveguides. Second, as shown in Fig.~\ref{fig:figScaling}(b), both approaches have a length $L = \alpha_1 N$, where $\alpha_1=2 L_\pi$ for the ProMMI and $L_\pi$ is the distance for one phase-degree of freedom. $L/N$ is higher for the MZM as this architecture must also include 50:50 couplers.
However, to keep the cross-talk specified in Eq.~(\ref{eq:maxTransfer}) below a maximum value, $\eta_{m',n'}^{\text{max}}$, the device length $L_z$ must be lower-bounded. With increasing $N$, the propagation constants $\beta_{m}$ become more and more closely spaced, reducing the denominator in Eq.~(\ref{eq:maxTransfer}). To counter this increase, $\kappa_{m',n'}^{(l)}$ must be commensurately reduced, thereby increasing the device length. As detailed in the Supplementary, by solving for $\kappa_{m',n'}^{(l)}$, the maximum cross-talk constraint $\eta_{m',n'}^{\text{max}}$ implies a device length  $L_z \geq \alpha_2 N^2$, where
\begin{equation*}
    \alpha_2 \approx \frac{\lambda } {n_\text{core} (1-(n_\text{clad}^2/n_\text{core}^2))}\sqrt{\frac{1- \left(\eta_{m',n'}^{\text{max}}\right)^2}{\left(\eta_{m',n'}^{\text{max}}\right)^2}}
\end{equation*}
The total device length $L_z$ is the greater of the $\mathcal{O}(N)$ and $\mathcal{O}(N^2)$ length requirements, yielding  $L_z=\text{max}(\alpha_1 N, \alpha_2 N^2)$.
Fig.~\ref{fig:figScaling}(b) shows this linear to quadratic cross-over  for various values of $\eta_{m',n'}^\text{max}$ and a nominal $L_\pi$ value used in state-of-the-art integrated lithium niobate modulators~\cite{Wang:18}. The cross-over happens when $\alpha_1 N= \alpha_2 N^2$, i.e., at $N=\alpha_1/\alpha_2$.

Some individual layers may need to be extended slightly to keep the total dielectric perturbation in layer $l$, $\sum_{(m,n)}\epsilon_{p,(m,n)}^{(l)}(\mathbf{r})$, below a practical maximum --- for example, the power supply voltage. In such instances, we can lower all perturbation amplitudes by the same factor $x_l$ and extend that layer's length by $1/x_l$.

Finally, the ProMMI architecture sharply curtails edge roughness losses due to reduced mode overlap. Since edge roughness commonly dominates the overall propagation loss rate $\Gamma$, we consider this advantage in detail. From coupled mode theory, the loss rate of the $m^\text{th}$ eigenmode due to coupling to radiative modes is given by 
\begin{equation}
    \Gamma_m \propto |\kappa_{m\beta}|^2 \rho(\beta)
\end{equation}
where $\rho(\beta)$ is the density of states attributed to radiative modes. The coupling coefficient $\kappa_{m\beta}$ is
\begin{equation}
    \kappa_{m\beta} = -\frac{i}{2} \frac{\omega^2}{c^2} \int A_m^*(x,y) \epsilon_R(x,y) A_\beta(x,y)\, dx \, dy
\end{equation}
where $\epsilon_R(x,y)$ is the permittivity perturbation causing radiative loss. For sidewall roughness, this perturbation is concentrated at the edges of the waveguide, i.e., $\epsilon_R(x) \approx \epsilon_R \delta(x-x_0)$, where the constant $\epsilon_R$ accounts for roughness and the Dirac delta function $\delta(x-x_0)$ centers on the sidewall. For such a perturbation, the coupling coefficient reduces to $\kappa_{m\beta} = -i(\omega^2/2c^2) \epsilon_R A_m^*(x_0)A_\beta(x_0)$, yielding a loss rate of
\begin{equation}
    \Gamma_m \propto \left(\frac{\omega^2}{2 c^2}\right)^2 \epsilon_R^2 A_m^2(x_0) |A_\beta(x_0)|^2 \rho(\beta)
\end{equation}
where, on average, $A_m^2(x_0) \propto 1/N$. Thus, the loss rate in an $N$-mode waveguide is $\sim\! N$ times smaller than in a single-mode waveguide. As shown in Fig.~\ref{fig:figScaling}(c), the ProMMI optical transmission relates to that of the MZM by $T_\text{ProMMI} \approx T_\text{MZM}^{1/N}$: i.e., losses are greatly reduced in the ProMMI when edge roughness dominates. Eliminating edge losses especially motivates ProMMI architectures in material platforms where smooth sidewalls are difficult to realize~\cite{Wang:18, Abel:19}. Other losses, possibly from active components, would then become dominant. To further reduce $\Gamma_m$, we can also lower $\epsilon_R$ by engineering waveguide edges~\cite{Lee:12}, which are challenging to implement in MZMs.

\section{Discussion}
Material platforms for the ProMMI should be selected based on function. Applications demanding high-speed programmability would favor electro-optic (EO) waveguide materials, such as thin-film lithium niobate~\cite{Wang:18} or barium titanate~\cite{Abel:19}. These materials have exceptionally low material losses~\cite{Zhang:17}, which the ProMMI architecture can reach by suppressing otherwise dominant edge roughness losses. Moreover, residual $\propto 1/N$ edge roughness losses are actually further suppressed in otherwise problematic trapezoidal waveguide profiles~\cite{Lee:12, Wang:18}. Applications with relaxed modulation speed requirements could combine passive waveguide layers with existing large-scale phase modulators. For example, the modes in a slab of silicon nitride (SiN)  covered in liquid crystals (LCs) could reach $\epsilon_{p,(m,n)}^{(l)}(\mathbf{r})/\epsilon_0 \sim 0.3$, combining excellent modulation contrast with the CMOS scalability of LC-on-silicon (LCOS) displays with millions of pixels. 

The ProMMI operating spectrum is determined by the phase-matching requirement as seen from Eq.~(\ref{eq:maxTransfer}). Though a lower index modulation $\Delta n$ reduces bandwidth, it reduces cross-talk due to partial phase-matching. This limitation is analogous to the bandwidth limitation due to modal dispersion in the couplers of MZMs. But whereas the MZM couplers are fixed, the ProMMI can be reprogrammed for different operating wavelengths -- e.g., by using a regular electrode array of $>N^2$ pixels. A CMOS driver backplane is well suited to the task as commercial LCOS systems have millions of electrodes across an area of $\sim\!1$ cm$^2$. Alternatively, holographic techniques could efficiently program the $N(N-1)$ coupling perturbations into photorefractive crystals (such as lithium niobate) with $N(N-1)$ projected laser fields~\cite{Heanue:94, Psaltis:95}.

The ProMMI architecture is remarkable in another aspect: it saturates the compactness bound of programmable optical transformations in packing $N^{2}$ degrees of freedom as tightly as the material allows. Each ProMMI modal degree of freedom requires an area of $A_\text{mode}= N L_\pi \lambda/ \sqrt{n_\text{core}^2-n_\text{clad}^2}$. Thus, assuming a lithium niobate ProMMI with an applied voltage of 0-1.5~V, a $1~\text{cm}^2$ chip area would enable full programmability of $N\sim 90$ modes. The universality of the ProMMI programmability naturally extends to other waveguide geometries  --- for example, a MM waveguide bend, described by an $N$-dimensional unitary evolution $U_{\text{bend}}$, can be incorporated into the ProMMI transformation as $U = \prod_{(m,n), l=1}^{N} T_{m,n}^{(l)} \rightarrow (\prod_{(m,n), l=n+1}^{N} T_{m,n}^{(l)})U_\text{bend}(\prod_{(m,n), l=1}^{n} T_{m,n}^{(l)})$.  

In conclusion, we introduced an architecture for programmable mode transformations that combines the best attributes of MZM and MMI constructions: (i) a constructive programming algorithm requiring the minimum $N^2$ control degrees of freedom; (ii) length scaling $\propto N$ for $N<\alpha_1/\alpha_2$ and otherwise $\propto N^2$; (iii) width scaling  $\propto N$ but without the “dead-space” between MZM waveguides; (iv) edge roughness loss rates reduced by $\mathcal{O}(1/N)$; (v) far easier waveguide fabrication as 50:50 couplers and bends are eliminated. Due to its large packing density, CMOS electronics are well suited to control the coupling perturbations. Holographic programming enables $N(N-1)$ control fields to control the $N(N-1)$ coupling perturbations. When combined with optical nonlinearities in the waveguide, the ProMMI architecture should be of great use in applications spanning deep learning~\cite{Shen:17, Wetzstein:20} to physics-based simulators~\cite{Christodoulides:03,Lederer:08,Segev:13}, to general-purpose quantum computing in photonic random walks~\cite{Lahini:18}. Moreover, the ProMMI saturates `control density': i.e., it is as compact as possible to fit $\mathcal{O}(N(N-1))$ control degrees of freedom.  This high control density, together with the potential for ultra-low optical loss and constructive programmability, will benefit applications ranging from LIDAR to virtual- and augmented- reality displays, to high-density telecom optical switches, machine learning accelerators, and optical quantum computing devices. We anticipate moreover that our approach extends beyond electromagnetic to other propagating bosonic fields, such as programmable multimode acoustic modes in a slab or magnons in 2D spin ensembles.

\vspace{10 pt}
\noindent \textbf{Funding}
Hugo Larocque acknowledges the support of the Natural Sciences and Engineering Research Council of Canada (NSERC), the MITRE Corporation Moonshot Program, the National Science Foundation (NSF, Award no. ECCS-1933556), and of the QISE-NET program of the NSF. D.E. acknowledges support from the Defense Advanced Research Projects Agency (DARPA) ONISQ program.
\vspace{10 pt}

\noindent \textbf{Acknowledgements}
The authors acknowledge Dr. Michael Fanto, Saumil Bandyopadhyay, Dr. Ryan Hamerly, and Prof. David A.B. Miller for fruitful discussions.

\vspace{10 pt}
\bibliography{prommiBib}

\end{document}


\title{Supplementary Document \\
Universal linear optics by programmable multimode interference}

\author{Hugo Larocque}
\affiliation{Research Laboratory of Electronics \& Department of Electrical Engineering and Computer Science, Massachusetts Institute of Technology, Cambridge, Massachusetts 02139, USA}

\author{Dirk Englund}
\affiliation{Research Laboratory of Electronics \& Department of Electrical Engineering and Computer Science, Massachusetts Institute of Technology, Cambridge, Massachusetts 02139, USA}
\email{hlarocqu@mit.edu}

\begin{abstract}
\end{abstract}

\maketitle

\section{Coupled Mode Theory}

\label{sec:CMT}

As discussed in the main text, we consider the evolution of an optical field propagating along the $z$ direction through the $l^\text{th}$ segment of an $N$-mode MMI waveguide perturbed by a structured permittivity profile $\epsilon^{(l)}_{p}(\mathbf{r})=\epsilon_A\epsilon_L(z)\epsilon_T(x,y)$. The propagation of the field is defined by the following $N$ sets of coupled differential equations:
\begin{equation}
    \label{eq:coupleModes}
   \frac{\partial{a_m}}{\partial z} e^{-i\beta_m z} = \frac{1}{\beta_m} \sum_n a_n \epsilon_{A}\epsilon_L(z)\kappa^{(l)}_{mn} e^{-i\beta_n z},
\end{equation}
where $a_m$ is the field's $m^\text{th}$ eigenmode expansion coefficient, $\beta_m=2\pi n_{\text{eff},m}/\lambda$ is the corresponding propagation constants expressed in terms of the mode's effective index $n_{\text{eff},m}$ and the optical wavelength $\lambda$. $\kappa_{mn}^{(l)}$ are the coupling constants defined as:
\begin{equation}
    \label{eq:coupling}
   \kappa^{(l)}_{mn} = -\frac{i}{2} \frac{\omega^2}{c^2} \int A_m^*(x,y) \epsilon_T (x,y) A_n(x,y) \, dx \, dy.
\end{equation}
where $\omega$ is the optical frequency, $c$ the velocity of light in free-space, and $A_m$ the eigenmode's transverse profile. We rely on these equations to construct permittivity perturbations that apply SU(2) transformations on pairs of modes supported by the waveguide. As in the case of SU(2) transformations in MZM meshes~\cite{Bogaerts:20}, we rely on perturbations defined over a propagation distance $\Delta z = L_\text{layer}$ that sequentially alter the relative phase and amplitudes of the modes. The length of each layer is divided equally into phase shifting and amplitude coupling perturbations.

\subsection{Phase shifting perturbations}

\begin{figure}[h!]
	\centering
	\includegraphics[width=\linewidth]{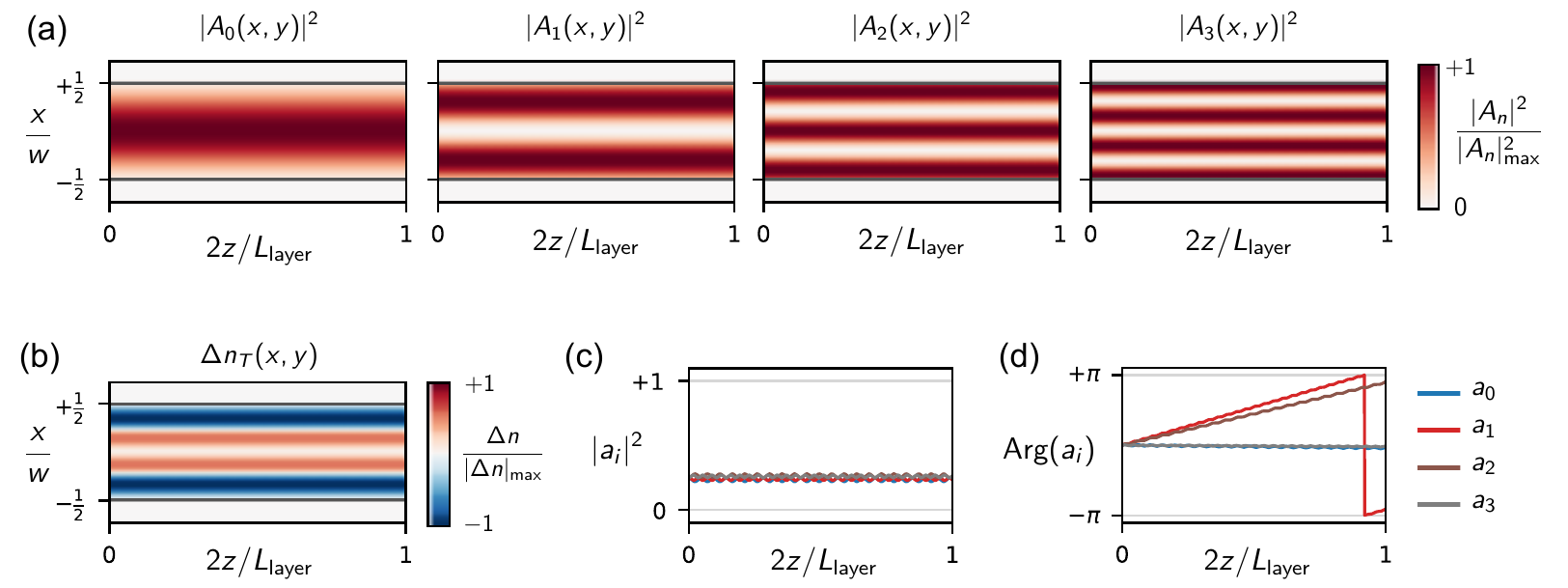}
	\caption{\textbf{Phase shifting with structured index perturbations.} (a) Eigenmode intensity profiles, $|A_i(x,y)|^2$, used to construct the phase shifting perturbations in a 4-mode waveguide. (b) Refractive index perturbation, $\Delta n_T(x,y)$, attributed to an $\epsilon_T(x,y)$ applying phase shifts of 0, -2.79, 2.79, and 0 to modes $A_{i=0,\dots,3}$, respectively. (c) Absolute square and (d) phase of $a_n$  for an input optical field $(a_0,a_1,a_2,a_3)^T=(1,1,1,1)^T/4$.}
	\label{fig:Phase}
\end{figure}

To impart phase shifts onto the modes of the waveguide, we rely on perturbations where $\epsilon_L(z)=1$, which modify the propagation constant of the $m^\text{th}$ eigenmode to $\tilde\beta^{(l)}_m$. Therefore, for such a perturbation defined over a length $ L_\text{layer}/2$ , the $m^\text{th}$ eigenmode experiences a phase shift of $\phi_m^{(l)} = (\tilde\beta^{(l)}_m-\beta^{(l)}_m)L_\text{layer}/2$.
We construct $\epsilon_T(x,y)$ as a weighted sum of the intensity profile of each mode, i.e, $\epsilon_T(x,y)=\sum_n c_n |A_n(x,y)|^2$. Each term will predominantly affect the propagation of its respective mode, hence enabling control over the phase of all modes supported by the waveguide. Figure~\ref{fig:Phase} shows the $\epsilon_T(x,y=0)$ implemented in layer $l=1$ of the $N=4$-mode slab waveguide considered in the main text. Namely, the considered waveguide has a core index $n_\text{core}=2.2886$, a cladding index $n_\text{clad}=1.44$, and a width of $w=1.525$~\textmu m. In Fig.~\ref{fig:Phase}(a), we provide the $|A_n(x,y)|^2=|A_n(x)|^2$ used to construct the $\epsilon_T(x,y)$ provided in Fig.~\ref{fig:Phase}(b). We then consider an input field with an optical wavelength of 1.55~\textmu m formed by an equal superposition of the waveguide's eigenmodes propagating through this perturbation. The corresponding relative absolute squares and phases of the eigenmodes are plotted in Fig.~\ref{fig:Phase}(c),(d) respectively.
As required, the modes' relative amplitudes remain mostly constant upon propagation whereas their phases linearly increase to their target values after a distance of $L_\text{layer}/2$. Small oscillations are observed in $a_i$ due to mode coupling enabled by the perturbations. However, as quantified by Eq.~(6) of the main text, the amplitude of these oscillations is constricted by the phase mismatch between the modes and the magnitude of the perturbations. These oscillations can also be further reduced by decreasing the amplitude of $\epsilon_T$, albeit at the expense of increasing $L_\text{layer}$.

\subsection{Coupling perturbations}

Equation~(\ref{eq:coupleModes}) is characterised by a phase mismatch $(\beta_m - \beta_n)z$. Therefore, to enable arbitrary conversions between a pair of modes, $\epsilon_L(z)$ should have a period of $\Lambda_{mn}=2\pi/|\beta_m-\beta_n|$ in the $z$ direction, thereby enabling perfect phase matching between the coupled modes~\cite{Yariv:73}. The strength of this interaction is quantified by the coupling constants from Eq.~(\ref{eq:coupling}) and is determined by the transverse profile of the perturbation. In order to attain reasonable coupling strengths, we rely on perturbations defined by the product of the coupled spatial modes’ profiles~\cite{Tseng:06, Ohana:14}, i.e., $\epsilon_T(x,y) = A_n^*(x,y) A_m(x,y)$.  

\begin{figure}[h!]
	\centering
	\includegraphics[width=\linewidth]{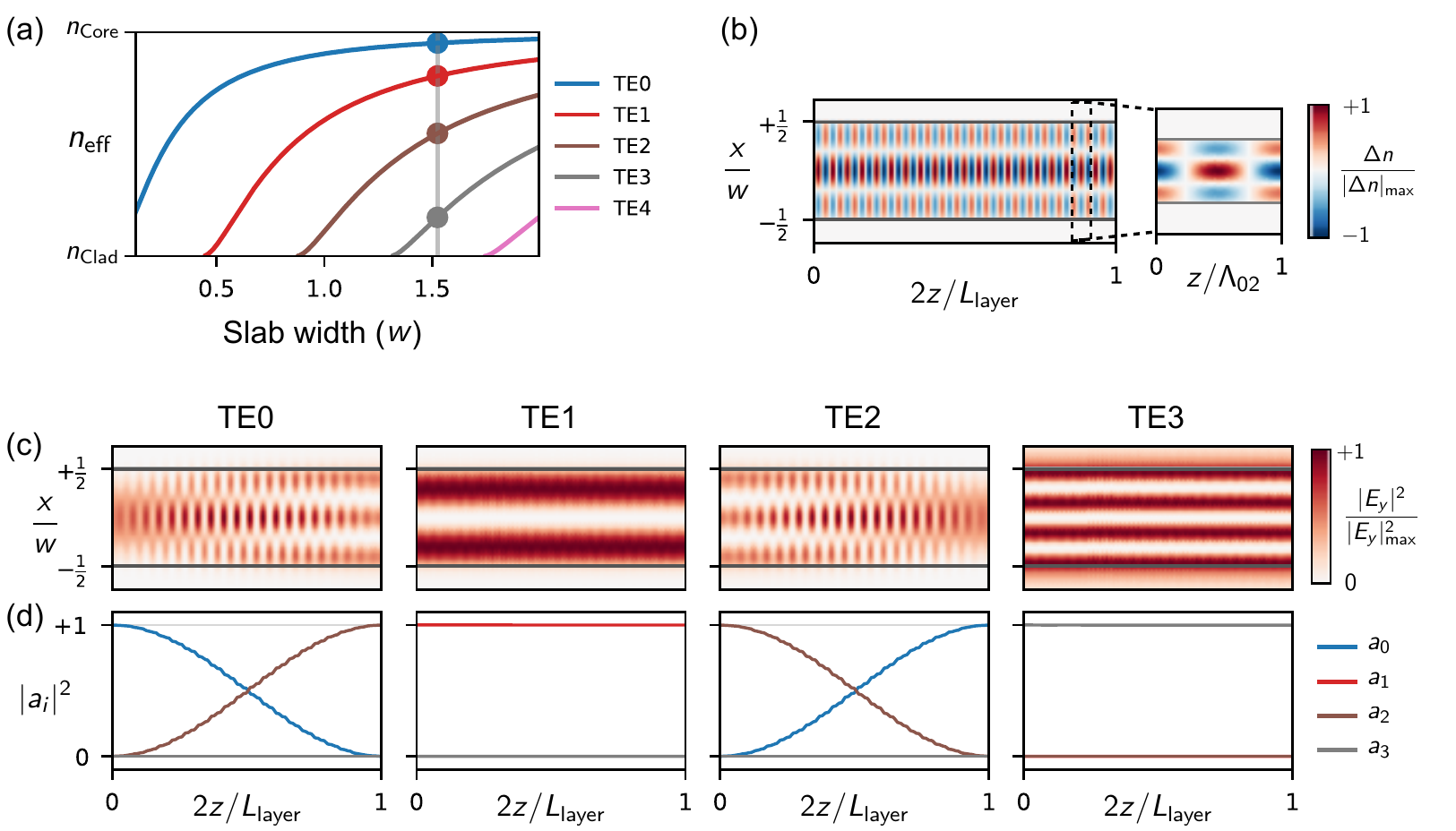}
	\caption{\textbf{Mode coupling with structured index perturbations.} (a) TE mode indices, $n_\text{eff}$, of a slab waveguide with a core and a cladding whose refractive indices are given by $n_\text{Core}=2.2886$ and $n_\text{Clad}=1.44$, respectively. The gray vertical line corresponds to the slab width considered in the main text and in the rest of the figure. Its intersection with each index curve provides the mode indices of the waveguide. (b) Structured index perturbation designed to couple the TE0 and TE2 modes of the slab. The period of the perturbation is given by $\Lambda_{02}=\lambda/(n_\text{eff,0}-n_\text{eff,2})$, where $\lambda$ is the optical wavelength, such as to enable a phase-matched interaction between the two modes. (c) Evolution of the optical field, $E_y$, upon propagating through a waveguide perturbed by the index profile shown in (b) for input profiles, $E_y(z=0)$, attributed to the slab's eigenmodes. (d) Absolute square of the eigenmode expansion coefficients for the fields shown in (c). The TE0 and TE2 modes are coupled due to being phase matched by the index profile, whereas any interaction involving the TE1 and TE3 modes are mismatched, thereby leaving them largely unperturbed as they propagate through the perturbed waveguide.}
	\label{fig:figConcept}
\end{figure}

In Fig.~\ref{fig:figConcept}, we consider an example of such a perturbation in the waveguide from Fig.~\ref{fig:Phase}. Figure~\ref{fig:figConcept}(a) plots the modes' effective indices as a function of the slab's width $w$. We then use the effective indices for the considered waveguide width to determine the period of $\epsilon_L(z)$ enabling mode coupling between the TE0 and TE2 modes of the slab. With this period, we construct the perturbation shown in Fig.~\ref{fig:figConcept}(b). The propagation of each eigenmode through this perturbation is then considered. The evolution of the corresponding $y$ component of the optical field, $E_y(\mathbf{r})$, and its eigenmode expansion coefficients are plotted in Fig.~\ref{fig:figConcept}(c),(d) respectively. As in the case of the phase shifting perturbations, small oscillations in the expansion coefficients occur due to the effects of partial phase matching and can be suppressed by lowering the magnitude of $\epsilon_p^{(l)}$ and by increasing $L_\text{layer}$.

\section{SU(N) Construction Algorithm}

The perturbations defined in Section~\ref{sec:CMT} allow the construction of any SU(2) transformations between a pair of modes. Namely, a phase shifting perturbation followed by a coupling perturbation between modes $m,n$ perform the following transformation on the amplitude coefficients $(a_m,a_n)^T$:
\begin{equation}
    \label{eq:su2}
    T^{(l)}_{mn} = 
    \begin{pmatrix}
        e^{-i\phi} \cos{\theta} & -i \sin{\theta} \\
       -i e^{-i\phi}\sin{\theta} & \cos{\theta} \\
    \end{pmatrix},
\end{equation}
where $\theta= \kappa^{(l)}_{mn}L_\text{layer}/2$ is set by the mode coupling and $\phi = ((\tilde\beta^{(l)}_m-\tilde\beta^{(l)}_n)-(\beta^{(l)}_m-\beta^{(l)}_n))L_\text{layer}/2$ by the phase shifting perturbations. Equation~(\ref{eq:su2}) consists of the fundamental building block for constructing arbitrary SU(N) transformations. We rely on the decomposition algorithm presented in~\cite{Clements:16} to find the $T^{(l)}_{mn}$ matrices whose product yields a target $N$-dimensional transformation, $U$. The algorithm involves sequentially finding $\phi, \theta$ parameters that null out elements of $T^{(l)}_{mn} U$ or $U (T^{(l)}_{mn})^{-1}$. However, differences in the formulation of the $T^{(l)}_{mn}$ used in this work and the one in~\cite{Clements:16} will carry over into the $\phi, \theta$ obtained in the two approaches. This decomposition will implement the unitary $U$ up to a phase, which must be corrected with a phase shifting perturbation at the waveguide's output.

Equation~(\ref{eq:su2}), implies that only one degree of freedom (DOF), $\phi$, is required to set the relative phase between a pair of modes. However, as mentioned in Section~\ref{sec:CMT}, we rely on $N$ DOFs, $c_n$, to construct an $\epsilon_T(x,y)$ imparting phase shifts on $N$ modes. This particular choice was made to enable full control over the phase of the modes, and not only pair-wise relative phases, in order to enable convenient parallels between the phase shifts used in~\cite{Clements:16}. Future iterations of our methods could include a reduction of DOFs down to the minimal amount required for arbitrary relative phases between pairs of modes

\section{Hardware Error Correction}

Refractive index perturbations that do not provide phase matched interactions between pairs of modes are one of the main sources of anticipated errors in our ProMMI architecture. This type of error specifically occurs when $\Lambda_{mn} \neq 2\pi/|\beta_{m}-\beta_{n}|$, where $\Lambda_{mn}$ is the period of the perturbation coupling modes $m,n$ with propagation constants $\beta_m,\beta_n$, respectively. These discrepancies modify the general SU(2) transformation provided in Eq.~(8) of the main text to 
\begin{equation}
\label{eq:errorMat}
\begin{pmatrix}
	(\cos{(\sqrt{\delta^2 + {(\kappa_{mn}^{(l)})}^2} z)} - i \frac{\delta}{\sqrt{\delta^2 + {(\kappa_{mn}^{(l)})}^2}} \sin{(\sqrt{\delta^2 + {(\kappa_{mn}^{(l)})}^2} z)}) e^{i (\delta z-\phi)} & -i \frac{{\kappa_{mn}^{(l)}}}{\sqrt{\delta^2 + {(\kappa_{mn}^{(l)})}^2}} \sin(\sqrt{\delta^2 + {(\kappa_{mn}^{(l)})}^2} z)e^{-i\delta z}\\
	-i \frac{\kappa_{mn}^{(l)}}{\sqrt{\delta^2 + {(\kappa_{mn}^{(l)})}^2}} \sin(\sqrt{\delta^2 + {(\kappa_{mn}^{(l)})}^2} z)e^{i (\delta z-\phi)} & (\cos{(\sqrt{\delta^2 + {(\kappa_{mn}^{(l)})}^2} z)} + i \frac{\delta}{\sqrt{\delta^2 + {(\kappa_{mn}^{(l)})}^2}} \sin{(\sqrt{\delta^2 + {(\kappa_{mn}^{(l)})}^2} z)})e^{-i\delta z},
\end{pmatrix}
\end{equation}
where
\begin{eqnarray*}
    z &=&{L_\text{layer}}/{2}, \\
	\delta &=& 	{(\beta_m-\beta_n-\beta_p)}/{2}, \\
	\beta_p &=& 2\pi/\Lambda_{mn}.
\end{eqnarray*}
As prescribed in~\cite{Bandyopadhyay:21}, these errors can be addressed by first  adjusting the coupling parameters of the above matrix such that the absolute value of its entries correspond to those of the error-free transformation. For the ProMMI, this means increasing the coupling strength ${\kappa_{mn}^{(l)}}$ to ${\kappa_{mn}^{(l)}}'$ such that
\begin{eqnarray*}
	\frac{({\kappa_{mn}^{(l)}}')^2}{\delta^2+({\kappa_{mn}^{(l)}}')^2} \sin^2{(\sqrt{\delta^2+({\kappa_{mn}^{(l)}}')^2}z)} =  \sin^2{({\kappa_{mn}^{(l)}} z)}.
\end{eqnarray*}
This equation is transcendental and cannot give us an exact closed form solution. We can nonetheless obtain an approximate one if we assume that $\delta, {\kappa_{mn}^{(l)}}' -{\kappa_{mn}^{(l)}} \ll {\kappa_{mn}^{(l)}}$. Let $({\kappa_{mn}^{(l)}}')^2=({\kappa_{mn}^{(l)}})^2 + \Delta \kappa^2$. Our transcendental equation becomes:
\begin{eqnarray*}
	\frac{({\kappa_{mn}^{(l)}})^2 + \Delta \kappa^2}{\delta^2+({\kappa_{mn}^{(l)}})^2 + \Delta \kappa^2} \sin^2{(\sqrt{\delta^2+({\kappa_{mn}^{(l)}})^2 + \Delta \kappa^2}z)} &=&  \sin^2{({\kappa_{mn}^{(l)}} z)}, \\
	\frac{1 + \Delta \kappa^2/({\kappa_{mn}^{(l)}})^2}{1+ (\delta^2 + \Delta \kappa^2)/({\kappa_{mn}^{(l)}})^2} \sin^2{\left({\kappa_{mn}^{(l)}} z \sqrt{1 + \frac{\delta^2 + \Delta \kappa^2}{({\kappa_{mn}^{(l)}})^2}}\right)} &=&  \sin^2{({\kappa_{mn}^{(l)}} z)}.
\end{eqnarray*}
For small errors, we expand the above equation to first order in $(\delta^2 + \Delta \kappa^2)/({\kappa_{mn}^{(l)}})^2$, hence
\begin{eqnarray*}
	\left(1+\frac{\Delta \kappa^2}{({\kappa_{mn}^{(l)}})^2}\right) \sin^2{({\kappa_{mn}^{(l)}} z)}+ \left(\frac{\delta^2+\Delta \kappa^2}{({\kappa_{mn}^{(l)}})^2}\right) \left(1+\frac{\Delta \kappa^2}{({\kappa_{mn}^{(l)}})^2}\right) ({\kappa_{mn}^{(l)}} z \cos({\kappa_{mn}^{(l)}} z)\sin({\kappa_{mn}^{(l)}} z)-\sin^2({\kappa_{mn}^{(l)}} z))&=&  \sin^2{({\kappa_{mn}^{(l)}} z)}, \\
	\frac{\Delta \kappa^2}{({\kappa_{mn}^{(l)}})^2} \sin^2{({\kappa_{mn}^{(l)}} z)}+ \left(\frac{\delta^2+\Delta \kappa^2}{({\kappa_{mn}^{(l)}})^2}\right) \left(1+\frac{\Delta \kappa^2}{({\kappa_{mn}^{(l)}})^2}\right) ({\kappa_{mn}^{(l)}} z \cos({\kappa_{mn}^{(l)}} z)\sin({\kappa_{mn}^{(l)}} z)-\sin^2({\kappa_{mn}^{(l)}} z))&=&  0, \\
	{\Delta \kappa^2}+ ({\delta^2+\Delta \kappa^2}) \left(1+\frac{\Delta \kappa^2}{({\kappa_{mn}^{(l)}})^2}\right) ({\kappa_{mn}^{(l)}} z \cot({\kappa_{mn}^{(l)}} z)-1)&=&  0.
\end{eqnarray*}
Furthermore, let us neglect $\delta^2 \Delta \kappa^2$ and $\Delta \kappa^4$ since they are much smaller than $\delta^2$ and  $ \Delta \kappa^2$. Doing so reduces the above to
\begin{eqnarray*}
	{\Delta \kappa^2}+ \left({\delta^2+\Delta \kappa^2}\right) ({\kappa_{mn}^{(l)}} z \cot({\kappa_{mn}^{(l)}} z)-1)&=&  0, \\
	\left({\delta^2+\Delta \kappa^2}\right) ({\kappa_{mn}^{(l)}} z \cot({\kappa_{mn}^{(l)}} z)) - \delta^2&=&  0, \\
	\Delta \kappa^2 ({\kappa_{mn}^{(l)}} z \cot({\kappa_{mn}^{(l)}} z))&=&  \delta^2 (1 - {\kappa_{mn}^{(l)}} z \cot({\kappa_{mn}^{(l)}} z)), \\
	\Delta \kappa^2 &=&  \delta^2 \left(\frac{\tan({\kappa_{mn}^{(l)}} z)}{{\kappa_{mn}^{(l)}} z } - 1\right),
\end{eqnarray*}
which determines the incremental coupling required to correct for errors in the coupling region. This correction diverges for ${\kappa_{mn}^{(l)}} z = \pi/2$, which is attributed to a perturbation performing a complete mode conversion, which cannot occur under phase-mismatched conditions. Assuming that this correction can be made, then it also introduces extraneous phases $\xi_a$, $\xi_b$, $\xi_c$, $\xi_d$, to the elements of Eq.~(\ref{eq:errorMat}), which are given by:
\begin{eqnarray*}
    \xi_a &=& - \arctan \left(\frac{\delta}{\sqrt{\delta^2 + ({\kappa_{mn}^{(l)}})^2}} \tan \left(\sqrt{\delta^2 + ({\kappa_{mn}^{(l)}})^2} z\right)\right),\\
    \xi_b &=& 0,\\
    \xi_c &=& 0,\\
    \xi_d &=& \arctan \left(\frac{\delta}{\sqrt{\delta^2 + ({\kappa_{mn}^{(l)}})^2}} \tan \left(\sqrt{\delta^2 + ({\kappa_{mn}^{(l)}})^2} z\right)\right).
\end{eqnarray*}
As a result, the transformation can now be written out as
\begin{equation*}
    \begin{pmatrix}
        e^{-i\phi} e^{i (\xi_a+\delta z)} \cos({\kappa_{mn}^{(l)}} z) & -i e^{i (\xi_b-\delta z)} \sin({\kappa_{mn}^{(l)}} z) \\
       -i e^{-i\phi} e^{i (\xi_c+\delta z)} \sin({\kappa_{mn}^{(l)}} z) & e^{i (\xi_d - \delta z)} \cos({\kappa_{mn}^{(l)}} z)
    \end{pmatrix}
    =
    e^{-i\delta z}\begin{pmatrix}
       e^{i \xi_b} & 0 \\
       0 & e^{i \xi_d}
    \end{pmatrix}
    \begin{pmatrix}
        e^{-i\phi} e^{i(2 \delta z + \xi_a - \xi_b)}\cos({\kappa_{mn}^{(l)}} z) & -i \sin({\kappa_{mn}^{(l)}} z) \\
       -i e^{-i\phi} e^{i(2 \delta z + \xi_a - \xi_b)}\sin({\kappa_{mn}^{(l)}} z) & \cos({\kappa_{mn}^{(l)}} z)
    \end{pmatrix},
\end{equation*}
where the above matrix factorization is enabled by $\xi_a + \xi_d = \xi_b + \xi_c$ as prescribed by unitarity. Therefore, to correct for these phase errors, $\phi$ must be adjusted such that $\phi \rightarrow \phi + 2\delta z + \xi_a-\xi_b$, and extra phases of $\psi_1 = \delta z - \xi_b$ and $\psi_2 = \delta z - \xi_d$ must be applied on the first and second modes attributed to the transformation, respectively.

To relate the mismatch $\delta$ to fabrication errors in the device's geometry, e.g. the waveguide's width, consider the propagation constants of the $m^\text{th}$ TE mode of a slab waveguide in the limit of a large waveguide with $w$~\cite{Cooney:16}:
\begin{equation*}
	\beta_m = k_0 n_\text{core} \sqrt{1 - \left(\frac{m \pi}{w_e k_0 n_\text{core}}\right)^2}\approx k_0 n_\text{core} - \frac{1}{2 k_0 n_\text{core}} \left(\frac{m \pi}{w_e}\right)^2,
\end{equation*}
where $k_0=2\pi/\lambda$ and $w_e = w + \lambda/\pi\sqrt{n_\text{core}^2-n_\text{clad}^2}$. For a perturbation coupling modes $m,n$, the corresponding period in a waveguide with a width of $w + \Delta w$ is given by
\begin{eqnarray*}
	\Lambda + \Delta \Lambda 
	&=& \frac{2\pi}{\beta_m-\beta_n}\\
	&\approx& \frac{2\pi}{\left(k_0 n_\text{core} - \frac{1}{2 k_0 n_\text{core}} \left(\frac{m \pi}{w_e + \Delta w}\right)^2 \right) - \left(k_0 n_\text{core} - \frac{1}{2 k_0 n_\text{core}} \left(\frac{n \pi}{w_e + \Delta w}\right)^2 \right) }\\
	&=& \frac{4 k_0 n_\text{core}(w_e+\Delta w)^2}{\pi (n^2- m^2) }\\
	&\approx& \frac{4 k_0 n_\text{core}}{\pi (n^2- m^2) } \left(w_e^2+2 w_e\Delta w\right),
\end{eqnarray*}
where $\Lambda$ is the corresponding period in a waveguide with a width $w$. Hence, fractional period variations can be expressed in terms of fractional width variations:
\begin{equation*}
	\frac{\Delta \Lambda}{\Lambda} \approx \frac{2\Delta w}{w_e}.
\end{equation*}
We can further express these changes in width in terms of fractional phase mismatch, which is more naturally incorporated into the equations describing errors in the ProMMI:
\begin{equation*}
	\frac{\Delta w}{w_e} = \frac{1}{2}\frac{\Delta \Lambda}{\Lambda} = \frac{1}{2}\frac{\Delta \beta_p}{\beta_p} = \frac{\delta}{\beta_p}.
\end{equation*}
Given that our perturbations only couple modes $m$ and $m+N/2$, $\beta_p$ is roughly independent of the number of modes supported by the waveguide. As a result, the mismatch error $\delta$ affecting the circuit is reduced as we consider waveguides with an increasing number of modes, thereby mitigating errors that come with greater circuit sizes.

\begin{figure}[h!]
	\centering
	\includegraphics[width=\linewidth]{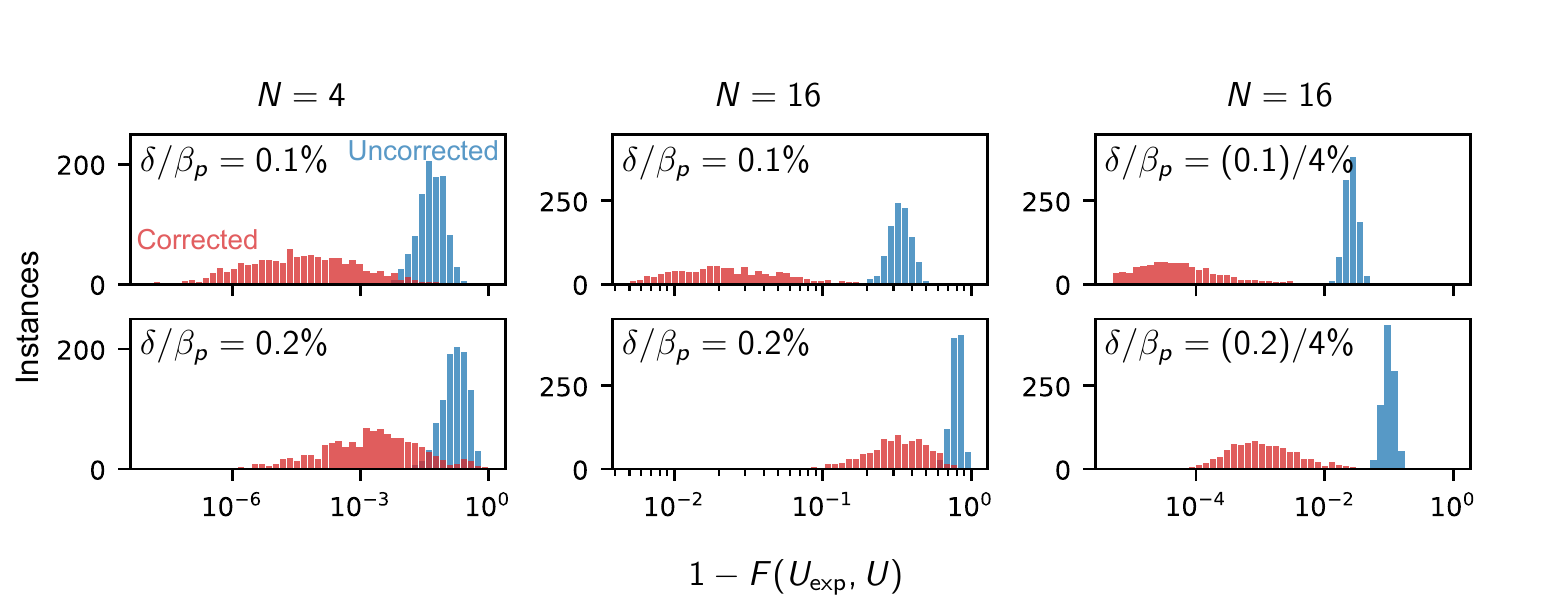}
	\caption{\textbf{ProMMI error correction and scaling.} Infidelity in the unitaries constructed from phase mismatched versions of Eq.~(\ref{eq:errorMat}) before and after correcting for imperfect phase matching. To emulate the effect of variations in the waveguide width from one layer to another, each SU(2) considered in the unitary's reconstruction was affected by a mismatch $\delta$ drawn from a normal distribution with a standard deviation of $\delta = \alpha \beta_p$. For the $N=4$ waveguide, $\alpha$ values of 0.1\% and 0.2\% correspond to width discrepancies of 1.5~nm and 3~nm, respectively. The corresponding width discrepancies for the $N=16$ mode waveguide are 4 times larger.}
	\label{fig:figError}
\end{figure}

Fig.~\ref{fig:figError} shows the matrix infidelity of 1000 randomly generated Haar unitary matrices implemented within the ProMMI architecture before and after using the above correction scheme. We assumed that the 2x2 matrices of Eq.~(\ref{eq:errorMat}), whose product yields the targeted SU(N) transformation when $\delta=0$, have $\delta$ values drawn from a zero-mean Gaussian distribution. This particular methodology emulates differences between the width of each layer in the ProMMI waveguide, which will differently affect mode perturbations based on their periods. Let us note that these results ignore errors attributed to modal cross-talk. $N=4$ and $N=16$ mode circuits are considered here. For the $N=16$ circuit, errors with the same mismatch $\delta$ as the $N=4$ circuit are considered. In addition, errors attributed to the same width variations are also considered, which, in the $N=16$ case, manifest as phase mismatches that are roughly 4 times smaller than in the $N=4$ circuits.

\section{Scaling Properties}

\subsection{Width}

\subsubsection{MZI Mesh}

The waveguides of an MZI mesh ought to be separated such as to have minimal coupling between the modes of each waveguide. Because the decay of the field outside the waveguide is exponential, let us quantify this overlap by approximating the mode of a waveguide by $\sqrt{1/\alpha}\exp{(- |x|/\alpha)}$, where $\alpha \sim \lambda/\pi \sqrt{n_\text{core}^2-n_\text{clad}^2}$. The overlap between two modes separated by $\Delta x$ therefore becomes:
\begin{eqnarray*}
	\kappa &=& \frac{\omega}{4} \frac{2\omega\mu_0}{k_0 n_\text{eff}} \epsilon_0 (n_\text{core}^2-n_\text{clad}^2) \int_{-\infty}^{+\infty} \frac{1}{\alpha}\exp{\left(-\frac{|x-\Delta x/2|}{\alpha}\right)}\exp{\left(-\frac{|x+\Delta x/2|}{\alpha}\right)} \\
	&=& \frac{\omega^2}{2c^2 k_0 n_\text{eff}} (n_\text{core}^2-n_\text{clad}^2)  \exp{\left(-\frac{\Delta x}{\alpha}\right)} \left(1 + \frac{\Delta x}{\alpha}\right)\\
	&=& k_0 \frac{(n_\text{core}^2-n_\text{clad}^2)}{2n_\text{eff}}  \exp{\left(-\frac{\Delta x}{\alpha}\right)} \left(1 + \frac{\Delta x}{\alpha}\right).
\end{eqnarray*}
Therefore, the distance between two waveguides with an optical overlap $\kappa$ is given by
\begin{eqnarray*}
	\Delta x &=& - \left({1+W_{-1}\left(\frac{2\kappa n_\text{eff}}{e k_0 (n_\text{core}^2-n_\text{clad}^2)}\right)}\right){\alpha} \\
	&\sim& -  \left(1+W_{-1}\left(\frac{2\kappa n_\text{eff}}{e k_0 (n_\text{core}^2-n_\text{clad}^2)}\right)\right) \frac{\lambda}{\pi \sqrt{n_\text{core}^2-n_\text{clad}^2}}
\end{eqnarray*}
where $W_{-1}(z)$ is the Lambert or product log function of order $-1$. To avoid having a cross-talk of $\eta$ due to undesired coupling between the waveguides, $\kappa$ must verify
\begin{eqnarray*}
	\sin^2{(\kappa (N L_\text{layer}))} < \eta \\
	{(\kappa (N L_\text{layer}))}^2 < \eta \\
	\kappa < \frac{\eta^{1/2}}{N L_\text{layer}}
\end{eqnarray*}
where $N$ is the number of modes, $L_\text{layer}$ is the length of a layer in the MZM, and where we assumed that $\eta$ was small. This implies that $\Delta x$ must satisfy:
\begin{eqnarray}
	\Delta x &>&  -  \left(1+W_{-1}\left(\frac{2 n_\text{eff} \eta^{1/2} }{e k_0 (n_\text{core}^2-n_\text{clad}^2)N L_\text{layer}}\right)\right) \frac{\lambda}{\pi \sqrt{n_\text{core}^2-n_\text{clad}^2}}
\end{eqnarray}
We can obtain an estimate of $\Delta x$ by subbing in values associated to state-of-the-art lithium niobate photonics~\cite{Wang:18}. Let us take $n_\text{core}=2.21$, $n_\text{clad}=1.444$, $n_\text{eff}=1.86$. If we want less than 1\% cross-talk, then $\eta^{1/2}=0.1$. Let $L_\text{layer}=4~\text{cm}$. Finally, let us assume that we are operating at a wavelength of 1550 nm. As displayed in Fig.~4a of the main text, these values give us a $\Delta x$ values of roughly $17\alpha-22\alpha$ for $N$ ranging from 1 to 100.

Let us note that though thermo-optic phase shifters enable a lower $L_\text{layer}$ value, thermal cross-talk constrains $\Delta x$ to much larger values, which are typically around $10$~\textmu m. In units of $\alpha$, this corresponds to roughly 33.

\subsubsection{ProMMI}

The number of modes in a multimode slab waveguide asymptotically scales with $2 w\sqrt{n_\text{core}^2-n_\text{clad}^2}/\lambda$, where $w$ is the width of the waveguide. Therefore, as plotted in Fig.~4a of the main text, the width of an MMI waveguide scales with
\begin{equation}
	w = \frac{N\pi}{2}\frac{\lambda}{\pi \sqrt{n_\text{core}^2-n_\text{clad}^2}}
\end{equation}

\subsection{Length}

As discussed in Section~\ref{sec:CMT}, structuring the spatial profile of a refractive index perturbation, $\Delta n(x,z)$, into the interference pattern attributed to modes $m$ and $n$ enables them to couple. In particular, for perturbations with a maximum index shift of $\text{max}(\Delta n(x,z)) = \Delta n_\text{max}$, the coupling rate between these modes, $\kappa_{mn}^{(l)}$, becomes comparable to the phase shift per unit length, $\Delta \phi / \Delta z$, achievable within single mode integrated phase shifters experiencing an index shift of $\Delta n_\text{max}$. In practice, these phase shifters are often designed to have a length enabling a $\pi$ phase shift, i.e. $L = L_\pi=\lambda / 2 \Delta n_\text{max}$. Therefore, for practical considerations, let us assume that the length of each layer in the ProMMI device is given by $L_\text{layer} = 2 L_\pi$, where a length of $L_\pi$ is allocated to the perturbation that introduces a relative phase between the waveguide's modes and the remaining distance is attributed to the perturbation coupling the modes. Let us note that this configuration also assumes that the ProMMI's index modulation can reach the maximum index shift of the sum of individual coupling perturbations overlaid on the multimode waveguide.

Based on the above construction, it follows that the length of an $N$-mode ProMMI device scales with $2N L_\pi$. For MZI meshes, length scales with $2N L_\pi (1 + L_\text{50:50}/L_\pi)$, where $L_\text{50:50}$ is the length of the 50:50 splitting elements used in the mesh. These elements could for instance consist of MMI splitters or directional couplers. Low-voltage state-of-the art EO phase shifters attributed to lithium niobate platforms~\cite{Wang:18} have a length 2~cm, hence a $L_\text{50:50}/L_\pi \approx 10^{-3}-10^{-2}$ as reflected in Fig.~4b of the main text.

\subsubsection{Linear to Quadratic Crossover}

As discussed in the main text, perturbations $\epsilon^{(l)}_{p,(m,n)}(\mathbf{r})$ in principle allow coupling between other pairs of modes $m',n'$ as indicated by the maximum transfer amplitude
\begin{eqnarray}
    \label{eq:maxTransfer}
    \eta_{m'n'} &=& \frac{2 |\kappa^{(l)}_{m'n'}|}{\sqrt{4 |\kappa^{(l)}_{m'n'}|^2+(\beta_{mn} - \beta_{m'n'})^2}}\\ \nonumber
    &=& \frac{1}{\sqrt{1+\left(\frac{\beta_{mn} - \beta_{m'n'}}{2|\kappa^{(l)}_{m'n'}|}\right)^2 }},
\end{eqnarray}
where $\beta_{mn}=\beta_m-\beta_n$, $\beta_{m'n'}=\beta_{m'}-\beta_{n'}$. These interactions are suppressed when 
\begin{equation*}
    |\kappa^{(l)}_{m'n'}|/|(\beta_{mn})-(\beta_{m'n'})| \ll 1,
\end{equation*}
which can be achieved in practice by lowering the amplitude of the perturbations. To quantitatively relate this constraint to a tolerable amount of error bounded by $\eta_{m'n'}$, we express $|\kappa^{(l)}_{m'n'}|$ in terms of the maximum transfer amplitude
\begin{equation*}
	|\kappa^{(l)}_{m'n'}| = \frac{|\beta_{mn}-\beta_{m'n'}|}{2} \sqrt{\frac{\eta^2_{m'n'}}{1- \eta^2_{m'n'}}}.
\end{equation*}
Each layer in the ProMMI needs to be able to perform full mode conversion, i.e. $\kappa^{(l)}_{mn}L_\text{layer}/2 = \pi/2$.  The coupling constant $\kappa^{(l)}_{m'n'}=\eta_{\kappa'}\kappa^{(l)}_{mn}$, where $\eta_{\kappa'}<1$ is a scaling factor that accounts for the mode overlap reduction in $\kappa^{(l)}_{m'n'}$ in the presence of $\epsilon^{(l)}_{p,(m,n)}(\mathbf{r})$. The layer length that suppresses errors down to $\eta_{m'n'}$ therefore becomes:
\begin{equation*}
	L_\text{layer} = \frac{2\pi \eta_{\kappa'}} {|\beta_{mn}-\beta_{m'n'}|}\sqrt{\frac{1- \eta^2_{m'n'}}{\eta^2_{m'n'}}}.
\end{equation*}
To express this length in terms of mode number, $N$, we rely on the asymptotic expression of the modes' propagation constants~\cite{Cooney:16}:
\begin{equation*}
	\beta_n = k_0 n_\text{core} \sqrt{1-\frac{n^2 (1-(n_\text{clad}/n_\text{core})^2)}{(N-\tfrac{1}{2}+\tfrac{2}{\pi})^2}} \approx k_0 n_\text{core} \left({1-\frac{1}{2}\frac{n^2 (1-(n_\text{clad}/n_\text{core})^2)}{(N-\tfrac{1}{2}+\tfrac{2}{\pi})^2}}\right)
\end{equation*}
where the approximation made above holds for small $n$. In this limit, the difference $\beta_{mn}$ is 
\begin{equation*}
	\beta_{mn} \approx \frac{k_0 n_\text{core}}{2} \frac{1}{(N-\tfrac{1}{2}+\tfrac{2}{\pi})^2} \left(1-\frac{n_\text{clad}^2}{n_\text{core}^2}\right)(n^2-m^2).
\end{equation*}
The ProMMI relies on the coupling between modes separated by roughly $N/2$ other ones. For such combinations, $\beta_{mn}$ becomes
\begin{eqnarray*}
	\beta_{m(m+N/2)} &\approx& \frac{k_0 n_\text{core}}{2}  \frac{1}{(N-\tfrac{1}{2}+\tfrac{2}{\pi})^2} \left(1-\frac{n_\text{clad}^2}{n_\text{core}^2}\right)((m+\tfrac{N}{2})^2-m^2)\\
	&\approx& \frac{k_0 n_\text{core}}{2}  \frac{1}{(N-\tfrac{1}{2}+\tfrac{2}{\pi})^2} \left(1-\frac{n_\text{clad}^2}{n_\text{core}^2}\right)\left(\frac{N^2}{4}+mN\right)\\
	&\approx& \frac{k_0 n_\text{core}}{8} \frac{N^2+4mN}{(N-\tfrac{1}{2}+\tfrac{2}{\pi})^2} \left(1-\frac{n_\text{clad}^2}{n_\text{core}^2}\right).
\end{eqnarray*}
The $\beta_{mn}$ and $\beta_{m'n'}$ that most significantly increase $L_\text{layer}$ are the ones where $m'=m+1$, hence
\begin{eqnarray*}
	\beta_{mn} - \beta_{m'n'} &=& \beta_{m(m+N/2)} - \beta_{(m+1)(m+1+N/2)} \\
	&=& \beta_{(m+1)(m+1+N/2)} - \beta_{m(m+N/2)} \\
	&=& \frac{k_0 n_\text{core}}{8} \left(1-\frac{n_\text{clad}^2}{n_\text{core}^2}\right) \left(\frac{N^2+4(m+1)N}{(N-\tfrac{1}{2}+\tfrac{2}{\pi})^2}- \frac{N^2+4mN}{(N-\tfrac{1}{2}+\tfrac{2}{\pi})^2} \right) \\
	&=& \frac{k_0 n_\text{core}}{8} \left(1-\frac{n_\text{clad}^2}{n_\text{core}^2}\right) \left(\frac{4N}{(N-\tfrac{1}{2}+\tfrac{2}{\pi})^2} \right) \\
	&=& \frac{k_0 n_\text{core} N}{2(N-\tfrac{1}{2}+\tfrac{2}{\pi})^2} \left(1-\frac{n_\text{clad}^2}{n_\text{core}^2}\right),
\end{eqnarray*}
Thereby setting $L_\text{layer}$ to 
\begin{eqnarray*}
	L_\text{layer} &=& \frac{2\pi } {k_0 n_\text{core} (1-(n_\text{clad}^2/n_\text{core}^2))} \frac{2(N-\tfrac{1}{2}+\tfrac{2}{\pi})^2}{N}\eta_{\kappa'}\sqrt{\frac{1- \eta^2_{m'n'}}{\eta^2_{m'n'}}}\\
	&\approx& \frac{2N\lambda } {n_\text{core} (1-(n_\text{clad}^2/n_\text{core}^2))}\eta_{\kappa'}\sqrt{\frac{1- \eta^2_{m'n'}}{\eta^2_{m'n'}}}.
\end{eqnarray*}
For small $N$, this distance is smaller than the distance $L_\pi$ required for one phase degree of freedom. In this regime, the ProMMI length thereby scales linearly with $N$. For larger $N$, this length is larger than $L_\pi$, and the scaling of the ProMMI length becomes quadratic. This occurs when $L_\text{layer}=2L_\pi$, i.e.
\begin{eqnarray*}
	2L_\pi &=& L_\text{layer} \\
	2L_\pi &=&  \frac{2N\lambda } {n_\text{core} (1-(n_\text{clad}^2/n_\text{core}^2))}\eta_{\kappa'}\sqrt{\frac{1- \eta^2_{m'n'}}{\eta^2_{m'n'}}} \\
	N &=& \frac{L_\pi n_\text{core}}{\lambda \eta_{\kappa'}}\left(1-\frac{n_\text{clad}^2}{n_\text{core}^2}\right)\sqrt{\frac{\eta^2_{m'n'}}{1- \eta^2_{m'n'}}},
\end{eqnarray*}
thereby prescribing the mode number at which length scaling becomes quadratic given an error $\eta_{m'n'}$. This crossover point depends on the material platform of the ProMMI and on allowable errors.

\subsection{Mode Density}

With the length and width estimates provided above, we can obtain the total area occupied by a ProMMI device. Let us first recall that the width of an $N$-mode slab waveguide scales as follows
\begin{equation}
	w = \frac{N \lambda}{2 \sqrt{n_\text{core}^2-n_\text{clad}^2}}.
\end{equation}
Furthermore, an $N$-mode ProMMI has a depth of $N$ layers of length $2L_\pi$. Therefore, the total area occupied by a ProMMI becomes
\begin{equation}
	A \sim w(2NL_\pi) \sim \frac{N^2 L_\pi \lambda}{ \sqrt{n_\text{core}^2-n_\text{clad}^2}}.
\end{equation}
As discussed in the main text, the above expression implies that a $90$-mode ProMMI device realized within state-of-the-art lithium niobate waveguides~\cite{Wang:18} should occupy an area of roughly 1 $\text{cm}^2$. However, larger-scale ProMMI devices are prone to errors attributed to modal cross-talk, which can be suppressed by relying on lower permittivity perturbations. Relying on such perturbations would in turn increase the $L_\pi$ value of the device, and hence reduce the total number of modes that can be supported within a fixed area.

\bibliography{prommiBibSupp}